\documentclass[manuscript]{aastex631}
\usepackage[utf8]{inputenc}
\usepackage[T1]{fontenc}
\usepackage{CJK}
 
\shorttitle{WFST SSOs}
\shortauthors{Wang et al.}
 
\graphicspath{{./}{figures/}}

\begin{document}
\begin{CJK*}{UTF8}{gbsn}

\title{A Heliocentric-orbiting Objects Processing System (HOPS) for the Wide Field Survey Telescope: Architecture, Processing Workflow, and Preliminary Results}

\correspondingauthor{Lulu Fan; Xu Kong}
\email{llfan@ustc.edu.cn; xkong@ustc.edu.cn}

\author[0009-0007-8133-5249]{Shao-Han Wang}
\altaffiliation{These authors contributed equally.}
\affiliation{Department of Astronomy, University of Science and Technology of China, Hefei 230026, China}
\affiliation{School of Astronomy and Space Science, University of Science and Technology of China, Hefei 230026, China}

\author[0000-0003-4122-6949]{Bing-Xue Fu}
\altaffiliation{These authors contributed equally.}
\affiliation{Department of Astronomy, University of Science and Technology of China, Hefei 230026, China}
\affiliation{School of Astronomy and Space Science, University of Science and Technology of China, Hefei 230026, China}

\author[0009-0008-4604-9674]{Jun-Qiang Lu}
\altaffiliation{These authors contributed equally.}
\affiliation{Department of Astronomy, University of Science and Technology of China, Hefei 230026, China}
\affiliation{School of Astronomy and Space Science, University of Science and Technology of China, Hefei 230026, China}

\author[0000-0003-4200-4432]{LuLu Fan}
\affiliation{Department of Astronomy, University of Science and Technology of China, Hefei 230026, China}
\affiliation{School of Astronomy and Space Science, University of Science and Technology of China, Hefei 230026, China}
\affiliation{Deep Space Exploration Laboratory, Hefei 230088, China}

\author[0000-0003-4721-6477]{Min-Xuan Cai}
\affiliation{Department of Astronomy, University of Science and Technology of China, Hefei 230026, China}
\affiliation{School of Astronomy and Space Science, University of Science and Technology of China, Hefei 230026, China}

\author{Ze-Lin Xu}
\affiliation{Department of Astronomy, University of Science and Technology of China, Hefei 230026, China}
\affiliation{School of Astronomy and Space Science, University of Science and Technology of China, Hefei 230026, China}

\author[0000-0002-7660-2273]{Xu Kong}
\affiliation{Department of Astronomy, University of Science and Technology of China, Hefei 230026, China}
\affiliation{School of Astronomy and Space Science, University of Science and Technology of China, Hefei 230026, China}
\affiliation{Deep Space Exploration Laboratory, Hefei 230088, China} 

\author{Haibin Zhao}
\affiliation{Purple Mountain Observatory, Chinese Academy of Sciences, Nanjing, 210023, China}
\affiliation{School of Astronomy and Space Science, University of Science and Technology of China, Hefei 230026, China}

\author{Bin Li}
\affiliation{Purple Mountain Observatory, Chinese Academy of Sciences, Nanjing, 210023, China}

\author[0009-0008-4858-1410]{Ya-Ting Liu}
\affiliation{School of Artificial Intelligence and Data Science, University of Science and Technology of China, Hefei, 230026, China}

\author[0000-0003-0694-8946]{Qing-feng Zhu}
\affiliation{Department of Astronomy, University of Science and Technology of China, Hefei 230026, China}
\affiliation{School of Astronomy and Space Science, University of Science and Technology of China, Hefei 230026, China}
\affiliation{Deep Space Exploration Laboratory, Hefei 230088, China} 

\author{Xu Zhou}
\affiliation{Key Laboratory of Optical Astronomy, National Astronomical Observatories, Chinese Academy of Sciences, Beijing 100101, China}

\author[0000-0002-3105-3821]{Zhen Wan}
\affiliation{Department of Astronomy, University of Science and Technology of China, Hefei 230026, China}
\affiliation{School of Astronomy and Space Science, University of Science and Technology of China, Hefei 230026, China}

\author{Jingquan Cheng}
\affiliation{Purple Mountain Observatory, Chinese Academy of Sciences, Nanjing, 210023, China}

\author[0000-0002-9092-0593]{Ji-an Jiang}
\affiliation{Department of Astronomy, University of Science and Technology of China, Hefei 230026, China}
\affiliation{National Astronomical Observatory of Japan, 2-21-1 Osawa, Mitaka, Tokyo 181-8588, Japan}

\author{Feng Li}
\affiliation{State Key Laboratory of Particle Detection and Electronics, University of Science and Technology of China, Hefei 230026, China}

\author{Ming Liang}
\affiliation{National Optical Astronomy Observatory (NSF's National Optical-Infrared Astronomy Research Laboratory), 950 N Cherry Ave., Tucson, AZ 85726, USA}

\author{Hao Liu}
\affiliation{State Key Laboratory of Particle Detection and Electronics, University of Science and Technology of China, Hefei 230026, China}

\author[0000-0003-1297-6142]{Wentao Luo}
\affiliation{Deep Space Exploration Laboratory, Hefei 230088, China}

\author{Zhen Lou}
\affiliation{Purple Mountain Observatory, Chinese Academy of Sciences, Nanjing, 210023, China}

\author{Hairen Wang}
\affiliation{Purple Mountain Observatory, Chinese Academy of Sciences, Nanjing, 210023, China}

\author{Jian Wang}
\affiliation{State Key Laboratory of Particle Detection and Electronics, University of Science and Technology of China, Hefei 230026, China}
\affiliation{Deep Space Exploration Laboratory, Hefei 230088, China}

\author{Tinggui Wang}
\affiliation{Department of Astronomy, University of Science and Technology of China, Hefei 230026, China}
\affiliation{School of Astronomy and Space Science, University of Science and Technology of China, Hefei 230026, China}
\affiliation{Deep Space Exploration Laboratory, Hefei 230088, China} 

\author[0000-0002-1935-8104]{Yongquan Xue}
\affiliation{Department of Astronomy, University of Science and Technology of China, Hefei 230026, China}
\affiliation{School of Astronomy and Space Science, University of Science and Technology of China, Hefei 230026, China}

\author{Hongfei Zhang}
\affiliation{State Key Laboratory of Particle Detection and Electronics, University of Science and Technology of China, Hefei 230026, China}

\author[0000-0002-1330-2329]{Wen Zhao}
\affiliation{Department of Astronomy, University of Science and Technology of China, Hefei 230026, China}
\affiliation{School of Astronomy and Space Science, University of Science and Technology of China, Hefei 230026, China}

\begin{abstract}

Wide-field surveys have markedly enhanced the discovery and study of solar system objects (SSOs). The 2.5-meter Wide Field Survey Telescope (WFST) represents the foremost facility dedicated to optical time-domain surveys in the northern hemisphere. To fully exploit WFST's capabilities for SSO detection, we have developed a heliocentric-orbiting objects processing system (HOPS) tailored for identifying these objects. This system integrates HelioLinC3D, an algorithm well suited for the WFST survey cadence, characterized by revisiting the same sky field twice on the majority of nights. In this paper, we outline the architecture and processing flow of our SSO processing system. The application of the system to the WFST pilot survey data collected between March and May 2024 demonstrates exceptional performance in terms of both temporal efficiency and completeness. A total of 658,489 observations encompassing 38,520 known asteroids have been documented, and 241 newly discovered asteroids have been assigned provisional designations. In particular, 27\% of these new discoveries were achieved using merely two observations per night on three nights. The preliminary results not only illuminate the effectiveness of integrating HelioLinC3D within the SSO processing system, but also emphasize the considerable potential contributions of WFST to the field of solar system science.

\end{abstract}


\section{Introduction}\label{sec:intro}

Solar system objects (SSOs) represent remnants from the initial stages of solar system formation. Early evolutionary processes like planetary migration left their imprints in the distributions of the orbital parameters, chemical compositions, and size distributions of the SSOs. The dynamical properties of the different populations have given rise to numerous models of planetary formation and evolution \citep{Tsiganis2005Nature, Morbidelli2005Nature,Gomes2005Nature, Walsh2011Nature,Raymond2017Sci}. They include small natural entities within the solar system, ranging in size from $\sim$1 meter to several hundred kilometers, such as near-Earth objects (NEOs), main belt asteroids (MBAs), trans-Neptunian objects (TNOs), and various other smaller groups of asteroids and comets \citep{Michel2015astebook,Ye_2019}. The study of these objects provides significant insight into the formation and evolution of planets, the potential dangers posed by near-Earth asteroids and comets \citep{Brown2013238}, and the origins of life and essential life-supporting materials, such as water \citep{Ye_2024}.

Wide-field surveys, including the Panoramic Survey Telescope and Rapid Response System \citetext{Pan-STARRS; \citealp{Denneau_2013}}, the Catalina Sky Survey \citep{Christensen2016}, and the Zwicky Transient Facility \citetext{ZTF; \citealp{Masci_2019}}, have facilitated a marked increase in the discovery of solar system objects. As of mid-2024, approximately 1.4 million minor planets, encompassing asteroids and 5,000 comets, have been cataloged \citep{Ye_2024}. Nonetheless, it is estimated that billions of small bodies exceeding 1 kilometers in size remain unobserved, particularly those icy bodies within the Oort Cloud \citep{deLeon2018}. Furthermore, characterization of these objects in terms of their physical and compositional properties lags significantly, due to the considerable efforts required and the expansive parameter space between identification and physical characterization \citep{Mahlke_2019}. The forthcoming LSST at the Vera C. Rubin Observatory is projected to yield millions of new discoveries, thereby substantially enhancing our understanding of the solar system \citep{Schwamb_2018}.

Wide Field Survey Telescope (WFST) is a time-domain survey facility located at the summit of Saishiteng Mountain, close to Lenghu Town, in Qinghai Province, dedicated to monitoring dynamic northern skies \citep{wang_2023}. With a field of view (FOV) of 6.5 square degrees, WFST achieves a 5$\sigma$ detection limit of 22.31, 23.42, 22.95, 22.43, 21.50, 23.61 magnitudes in the \textit{u}, \textit{g}, \textit{r}, \textit{i}, \textit{z} and \textit{w} bands, respectively \citep{Lei_2023}. The depth of the WFST, which is 2 magnitudes fainter than that of the ZTF in the \textit{gri} bands, provides significant advantages for the detection of SSOs. It is anticipated that the WFST will obtain comprehensive observations of known SSOs and identify fainter and smaller celestial bodies, thereby improving our understanding of the solar system \citep{wang_2023,ljq2025}. 

To maximize the potential for SSO detection in the WFST data, it is essential to develop an algorithm that complies with the WFST survey cadence. To balance survey efficiency with scientific objectives, WFST will avoid consecutive observations in a single band and instead conduct observations in two bands on the majority of nights \citep{wang_2023}. This strategy presents distinct challenges compared to traditional frameworks for SSO search algorithms and platforms. For example, algorithms such as the Pan-STARRS Moving Object Processing System \citetext{MOPS; \citealp{Denneau_2013}}, NEARBY \citep{Stefanut2018}, Umbrella \citep{Stanescu_2021}, and the ssos pipeline \citep{Mahlke_2019} typically require discovering objects by submitting single tracklets of three or more detections to the Minor Planet Center (MPC) within a single night. Furthermore, ZTF's Moving Object Discovery Engine \citetext{ZMODE; \citealp{Masci_2019}} is designed to work with ZTF's cadence linking objects with a minimum of four detections over four consecutive nights. Tracklet-less Heliocentric Orbit Recovery \citetext{THOR; \citealp{Moeyens_2021}} affords substantial advantages owing to its autonomy from intra-night tracklets and a non-requisite predefined cadence of observations within a search interval, although it has not been subjected to testing or optimization for NEOs.

HelioLinC, a multi-night linking algorithm \citep{holman2018}, is well suited for WFST because of its compatibility with the telescope's two-returns-per-night cadence and its ability to efficiently handle various types of SSOs. The algorithm was originally developed for LSST, which will visit observed fields twice each night and image the entire visible night sky approximately every three nights. The algorithm can link two detections within the same night to form a tracklet and can further connect inter-night tracklet associations to create ``tracks'', which are typically separated by 3-4 nights \citep{LSSTSbook2009, Ivezic_2019}. HelioLinC assumes a heliocentric distance and a rate of change to propagate tracklets to a common epoch, enabling the formation of high-confidence clusters. Additionally, it covers a wide variety of asteroid orbital populations from NEOs to TNOs, while operating at minimal computational expense, scaling as \begin{math}\mathcal{O}\end{math}$(NlogN)$ with the number of tracklets $N$. Notably, HelioLinC achieves completeness rates of around 90\% in labeled sets and exceeds 99\% in the MPC's Isolated Tracklet File (ITF) in initial results \citep{holman2018}. This linking algorithm has also been successfully applied to the Pan-STARRS1 survey \citep{kurlander2024}. Consequently, HelioLinC was chosen for the detection of moving objects in WFST, leveraging its proven effectiveness and efficiency.

This paper presents a heliocentric-orbiting objects processing system tailored for WFST, constructed upon the HelioLinC algorithm, herein referred to as HOPS. The paper is organized as follows. Section \ref{sec:wfst}  provides an overview of the WFST and the alert data generated by the real-time data reduction pipeline. Section \ref{sec:SSO Processing System} details the design and processing of our SSO processing system. In Section \ref{sec:results}, the processing system is applied to three months of data from the WFST pilot survey to evaluate its performance and present the preliminary results. Section \ref{sec:sum} gives a brief summary.

\section{WFST AND DATA PRODUCT} \label{sec:wfst}

WFST is a dedicated photometric survey facility developed collaboratively by the University of Science and Technology of China (USTC) and the Purple Mountain Observatory (PMO). Equipped with a 2.5-meter diameter primary mirror, an active optics system and a mosaic CCD camera featuring 0.73 gigapixels in the primary focal plane, WFST captures high-quality images over a 6.5-square-degree FOV \citep{wang_2023}. The telescope benefits from excellent observing conditions, including an average seeing of 0.75 arcsec and a background of 22.0 mag $\text{arcsec}^{-2}$ , ensuring high-quality data \citep{Deng2021353}. WFST is designed to survey the northern sky with unprecedented sensitivity, aiming to explore the variable universe and catch up the time-domain events, such as supernovae \citetext{SNe; \citealp{Humaokai_2023}}, tidal disruption events \citetext{TDEs; \citealp{LinZheyu_2022}}, optical counterparts of gravitational wave events \citep{Liuzhengyan_2023}, active galactic nuclei \citetext{AGN; \citealp{Su_2024}} variability and SSOs \citep{ljq2025}.

Upon obtaining each raw scientific exposure, the WFST real-time data reduction pipeline \citep{cai2025WFST}, which is developed based on the LSST pipeline \citep{lsststack}, processes the exposure instantaneously. The pipeline systematically performs the following stages: (1) Preprocessing: This involves overscan and bias subtraction, which is subsequently followed by flat-fielding and sky subtraction. (2) Image calibration: This step encompasses PSF modeling, source detection, and measurement, as well as astrometric and photometric calibrations. (3) Generation of calibrated science images: Instrumental signals are removed to create a calibrated science image together with its associated source catalog. (4) Template subtraction: A template image, derived from coadded images also produced by the WFST pipeline, is subtracted from the calibrated science image to produce a difference image. (5) Difference image analysis: Source detection and measurement are repeated on the difference image, resulting in a new source catalog and an alert product that identify sources regarded as ``variable''. The alert products are organized and published in \textit{avro} format, holding detection results for each variable source. The key contents of the alert data include the World Coordinate System (WCS) information and pixel coordinates, observation time, band, total and differential flux, dipole and trail measurements, and 30 $\times$ 30 pixel cutouts (approximately 10 $\times$ 10 arcseconds) centered on each alert position. These processes ensure that WFST can efficiently identify and characterize transient and variable sources, providing valuable data for follow-up studies \citep{cai2025WFST}.

\section{SSO Processing System For WFST} \label{sec:SSO Processing System}

\begin{figure}
    \centering
    \includegraphics[width=1\linewidth]{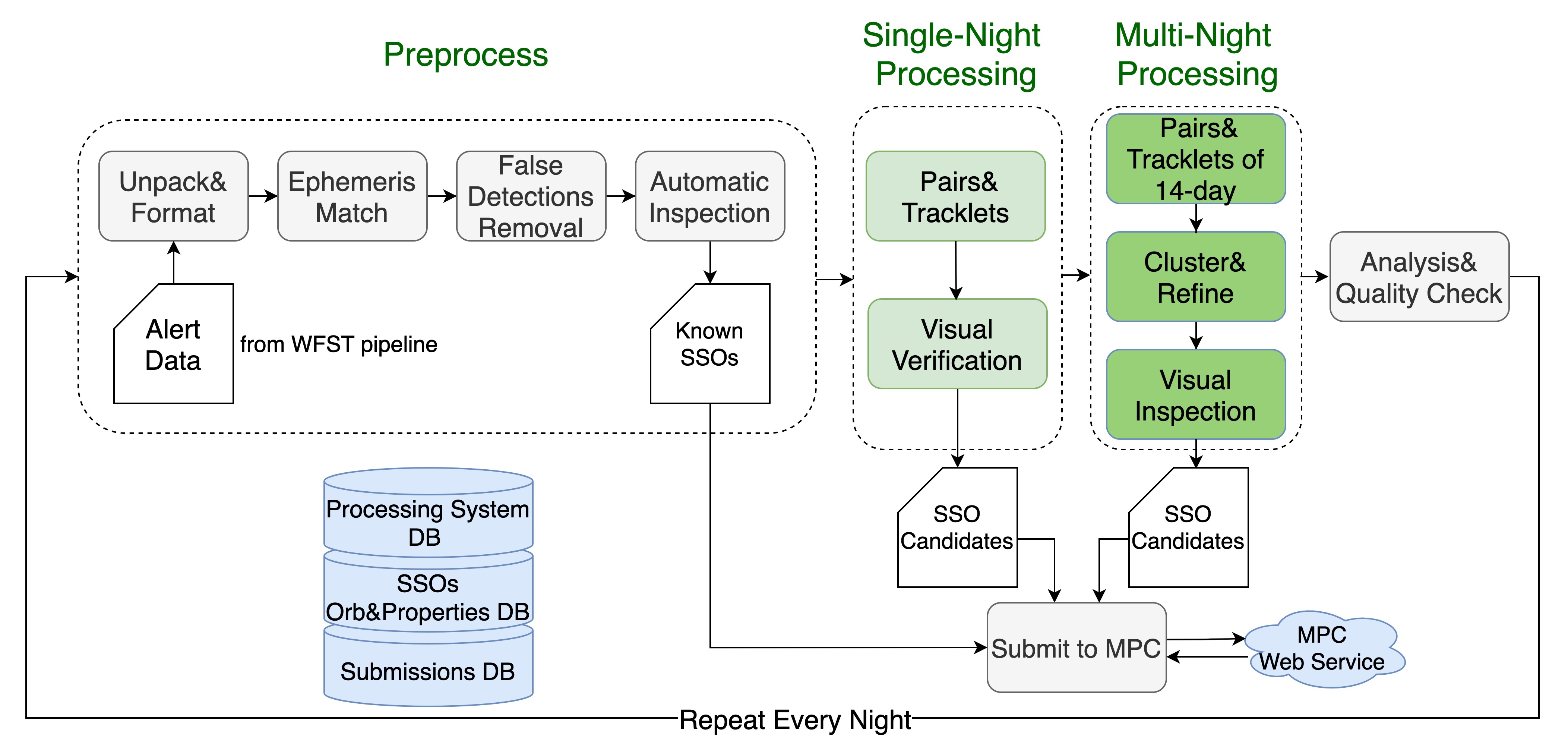}
    \caption{System architecture diagram of SSO processing system. The SSO processing system operates through these steps: (1) Preprocess: unpack and format alert data from the WFST pipeline, match with ephemerides, remove false detections, and automatically inspect known SSOs; (2) Single-Night Processing: identify pairs and tracklets, perform visual verification to confirm candidates; (3) Multi-Night Processing: analyze pairs and tracklets over a 14-day window, cluster and refine candidates, conduct final visual inspection; (4) Submission: Perform analysis and quality checks, submit known SSOs identified in step (1) and confirmed SSO candidates from steps (2) and (3) to the MPC through the web service; (5) Databases: maintain databases for processing system, orbital and property data, and submission records.}
    \label{fig:flowchart}
\end{figure}

Figure \ref{fig:flowchart} illustrates the processing flow of the SSO processing system. Upon receiving alert data from the WFST real-time data reduction pipeline, this processing system unpacks and formats the data for further processing. Key steps include: (1) Ephemeris Match: Calculating ephemerides to match known asteroids. (2) False Detection Removal: Employing machine learning and other methods to filter out false positives. (3) SSO search: Searching SSOs in both single-night and multi-night processing. Finally, after a thorough analysis and inspections, the processing system submits confirmed SSO detections to the Minor Planet Center (MPC). This section details each processing step.
\subsection{Preprocessing}

\subsubsection{Alert data unpack and format}

The SSO processing system ingests alert data in \textit{avro} format, unpacks the DIAsources (Difference Image Analysis Sources) from these files and formats them into \textit{csv} files. The corresponding cutouts are saved in \textit{fits} format. Each cutout includes a reference image, a science image, and a difference image, which are later used in the Convolutional Neural Network (CNN) for false detection removal. In addition, key measurements of each source, such as observation time, coordinates, and magnitudes, are extracted and computed for further analysis.

\subsubsection{Known asteroids match} \label{subsubsec:astcheck}

To match known asteroids, the processing system employs \texttt{astcheck}\footnote{\url{https://www.projectpluto.com/astcheck.htm}}, using the daily updated MPCORB dataset\footnote{\url{http://www.minorplanetcenter.org/iau/MPCORB.html}} from the MPC and JPL DASTCOM comet elements\footnote{\url{http://ssd.jpl.nasa.gov/dat/ELEMENTS.COMET}}. The \texttt{astcheck} calculates ephemerides of known SSOs，cross-matches them with all DIAsources based on both position and object motion. A threshold of 9 arcseconds is selected for matching due to the dimensions of the cutouts. Consequently, each source is annotated with its match results: marked as ``unknown'' if no match is found, or given the corresponding SSO designation along with the calculated distance, magnitude, and motion provided by \texttt{astcheck}.

\subsubsection{False detection removal}

A CNN model, \texttt{braai} \citep{Duev_2019}, is used to remove false detection and retain moving objects within our data. This established deep learning model was designed to classify real or bogus sources in ZTF. In our processing system, a real moving object should, as expected, appear as a point-like source in both the science and difference images but be absent in the reference image.

\begin{figure}
    \gridline
        {\fig{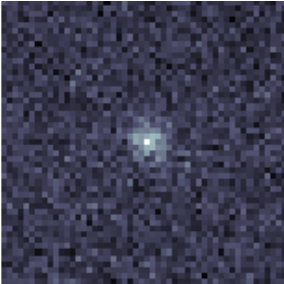}{0.2\textwidth}{}
         \fig{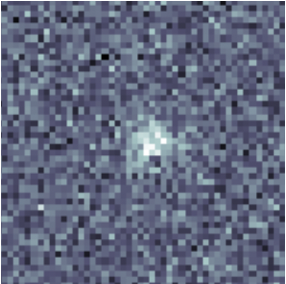}{0.2\textwidth}{}
         \fig{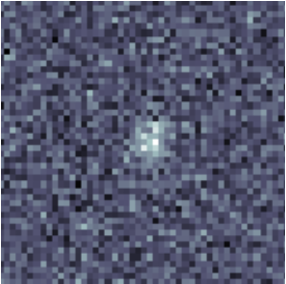}{0.2\textwidth}{}
         \fig{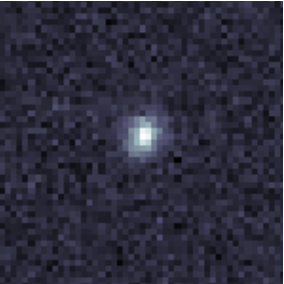}{0.2\textwidth}{}
        }
        {\fig{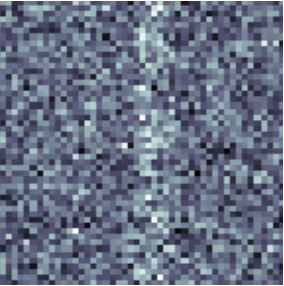}{0.2\textwidth}{}
         \fig{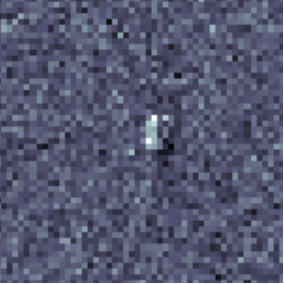}{0.2\textwidth}{}
         \fig{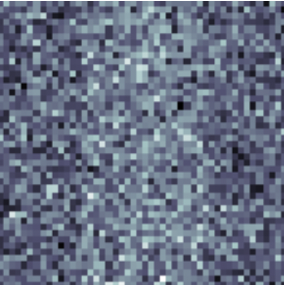}{0.2\textwidth}{}
         \fig{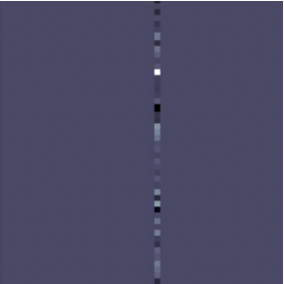}{0.2\textwidth}{}
        }
    \caption{Some example cutouts of true and false sources. The top row shows true sources and the bottom row shows false sources.
}
    \label{fig:MLexamples}
\end{figure}

During the early observation phase, the difference image process was not as reliable, so only the science image was used for classification. To train and evaluate the CNN model, we constructed a dataset comprising 4,272 true sources and 6,269 false sources. Figure \ref{fig:MLexamples} provides some examples of true and false sources. The sources were extracted as 50$\times$50 pixel cutouts and then manually labeled. These cutouts were divided into training and test sets in a 9:1 ratio. After 100 training epochs, the loss decreased to 0.0369, and the accuracy reached 0.9890, indicating that the model has achieved a stable and accurate state. To identify genuine asteroid candidates, we set criteria based on CNN scores: a science image cutout score of $\geq$ 0.55 and a reference image cutout score of $\leq$ 0.45. This approach eliminates the need for difference images during the early stages of observation.

\subsection{Searching and analysis} \label{subsec:Searching and analysis}

The open-source software \texttt{HelioLinC3D}\footnote{\url{https://github.com/lsst-dm/heliolinc2}\label{fn:heliolinc2}}, based on the HelioLinC algorithm \citep{holman2018}, has been incorporated into our processing system. The \texttt{HelioLinC3D} supports parallel execution, making it well suited to handle large-scale wide-field survey data sets. This integration enables our SSO processing system to operate in two distinct modes: single-night and multi-night processing. In single-night processing, the processing system analyzes data collected within a single night, generating tracklets with three or more detections. Meanwhile, in multi-night processing, the processing system utilizes a sliding 14-day window, encompassing data from the current night as well as the preceding 13 nights.

\subsubsection{Single-night processing}

For data collected over a single night, the processing system invokes \texttt{make\_tracklets}, which generates pairs and tracklets based on a configuration file specifying speed and arc length criteria. To validate a valid tracklet, \texttt{make\_tracklets} applies a least-squares fit, assuming that the object's motion follows a great circle with constant angular velocity\footref{fn:heliolinc2}. The results include pairs (comprising two detections) and tracklets (consisting of at least three detections). Upon completion of the single-night processing, a unique sequential identifier is assigned to each tracklet, with all detections belonging to the same tracklet sharing this identifier.
\subsubsection{Multi-night processing}

In multi-night processing, the processing system ingests data from a sliding 14-day window, incorporating observations from the preceding 13 nights and the current night. The \texttt{make\_tracklets} function is utilized to generate all possible pairs and tracklets. Subsequently, \texttt{heliolinc} converts these pairs and tracklets into three-dimensional positions and velocities, referred to as ``state vectors'', using predefined heliocentric distances and velocities. These state vectors are then transformed to a common reference time and clustered using the DBSCAN algorithm. To optimize efficiency, the process employs multiprocess parallel execution. Finally, the \texttt{link\_refine} function is executed to filter out redundant clusters, producing a de-duplicated, non-overlapping set of candidate discoveries with the highest likelihood of being real\footref{fn:heliolinc2}. All detections within the same cluster are assumed to be associated with the same asteroid. The cluster number (\textit{clusternum})  assigned to each detection is recorded as a result and stored in a \textit{csv} file.

In practice, \texttt{HelioLinC3D} demonstrates excellent scalability. It successfully identifies clusters with more than two detections per night for at least three nights within a 14-day window, as intended. Additionally, it can also cluster objects that have a minimum of six detections accumulated over at least two nights. For instance, an object with a minimum of three detections per night over two nights and one with two detections on the first night and four on the second, can both be effectively clustered.

\subsubsection{Analysis}

To assess the efficacy of search results, the highly regarded open-source tool \texttt{difi}\footnote{\url{https://github.com/moeyensj/difi}} is employed. This software utilizes \textit{diaSourceId} (a unique identifier for DIAsources), along with the names of the SSOs correlated by \texttt{astcheck} and the cluster number derived from the search results. The \texttt{difi} categorizes the clusters into pure, partial, mixed, and complete, thereby effectively demonstrating the relationship between the clusters and the associated known SSOs.

For single-night processing results, the processing system counts the number of \textit{findable} objects, those identified in the \texttt{astcheck} matched results and the number of \textit{found\_pure} (both found and pure) objects, both found and classified as pure in the search results. A minimum of 3 observations per tracklet is required for an object to be considered \textit{found\_pure}. For multi-night processing, the processing system similarly counts \textit{findable} and \textit{found\_pure} objects. The parameters \textit{linkage\_min\_obs} (minimum detections per night) and \textit{min\_linkage\_nights} (minimum number of nights) are utilized to accommodate various observation cadences. Typically, \textit{linkage\_min\_obs} is set at 2, while \textit{min\_linkage\_nights} is set at 3. Finally, the completeness of both single-night and multi-night processing is calculated as the ratio of \textit{found\_pure} to \textit{findable}, defined as the percentage of SSOs that are both findable and found in pure clusters. Moreover, \textit{pure\_unknown} results as SSO candidates are also extracted.

\subsection{Inspection and submission}

Before the results are submitted to the MPC, they must first undergo a series of thorough inspections. Known SSOs matched by the processing system are automatically inspected, while SSO candidates from both single-night and multi-night processing undergo visual verification. Observations believed to originate from real moving objects are retained only if they have at least two detections on the same night. Subsequently, formatted text files are generated, including batches of known SSOs and SSO candidates, complete with the necessary header lines. These files are then submitted to the MPC for further review and cataloging.

\subsubsection{Automatic inspections for known SSOs}\label{subsubsec:Automatic inspections for known SSOs}

When two or more detections are labeled with the same known SSO name, they are treated as a single specified object. Although the \texttt{astcheck} match and false detection removal processing greatly reduce the number of false detections, some false detections still persist. These are often due to poor photometry or morphology issues caused by image subtraction errors or gaps between detector chips. However, the high volume of detections that reach tens of thousands each night renders human visual inspection nearly unfeasible. Therefore, before submitting to the MPC, the processing system automatically verifies known SSOs based on predefined attributes such as band and magnitude.

A simplified approach is used to remove false detections for known SSOs. In a night, the magnitude variations of the same SSO should remain within a reasonable range, which can be verified by transforming the magnitudes from various bands to the $V$ band using color transformations. However, because WFST uses non-standard filters, the precise magnitude transformation relationships must be determined through photometric calibration. As a result, we are currently using existing magnitude transformation relations until the photometric calibration results are available. Specifically, for C and S type asteroids, we rely on color transformations from Johnson's $V$ band to the filters as an approximate solution. The observed magnitude transformed to the $V$ band $V_{tr}$ is calculated as follows: (1) $V_{tr} = g - 0.28$, (2) $ V_{tr} = r + 0.23$, (3) $V_{tr} = i + 0.39$, (4) $V_{tr} = z + 0.37$, and (5) $V_{tr} = u - 1.8$ \citep{Denneau_2013, jones2018181}. For detections of a known object, if the difference between the maximum and minimum $V_{tr}$ values exceeds 1.0, we exclude the detection that deviates the most from the median magnitude of all detections. Furthermore, any detection where the absolute difference between its $V_{tr}$ and the magnitude $V_{ast}$ provided by \texttt{astcheck} exceeds 1.0 is also excluded. This procedure helps to eliminate values that fall outside the acceptable range, thus reducing the impact of poor photometry or morphology issues.

\subsubsection{Visual inspections for SSO candidates}\label{subsubsec:Visual inspections for SSO candidates}

The visual inspection process serves as an essential concluding phase in the identification of SSO candidates. A direct approach includes the creation of animated sequences of images formatted in accordance with \textit{gif}. Within these \textit{gifs}, SSO candidates are expected to manifest almost consistent motion in a particular direction across multiple frames, thus making them identifiable against predominantly stationary background sources. In addition, an interactive website has been designed to show the \textit{ gifs} together with other pertinent information concerning the SSO candidates.

\subsubsection{Submitting procedure}  \label{subsubsec:Submitting procedure}

For SSO candidates that have successfully passed visual inspections, each observation is converted to the 80-column format stipulated by the MPC\footnote{\url{https://www.minorplanetcenter.net/iau/info/OpticalObs.html}}. To prevent duplicate submissions, the SSO candidates scheduled for submission undergo cross-matching with previously submitted entries in our database (see Section \ref{subsec:Database}). A candidate will not be resubmitted if it matches an existing entry within 10 seconds in time and 10 arcseconds in angular distance. Subsequently, header lines detailing observational information are appended to create a submission-ready text file, following MPC guidelines\footnote{\url{https://www.minorplanetcenter.net/iau/info/ObsDetails.html}}. The final submission is made using a script that uses the command-line method\footnote{\url{https://www.minorplanetcenter.net/iau/info/commandlinesubmissions.html}} to send observations to the MPC. The submission includes basic details of the batch, such as batch name, dates covered, and cluster numbers, which are documented for future reference and traceability. In total, there are three kinds of submissions: (1) single-night known asteroids, (2) single-night unknown asteroids, and (3) multi-night unknown asteroids. It is important to note that multi-night known asteroids are excluded from this list because they are consistently cataloged as single-night known asteroids on a nightly basis.

\subsection{Database\label{subsec:Database}}

We utilize \texttt{MongoDB}, a document-oriented, flexible, and high-performance NoSQL database, to efficiently manage the data of the processing system on the server. The \texttt{pymongo} module is employed within the processing system to handle all database interactions. In the processing system, a submission database is constructed to record automated acknowledgments from the MPC, each containing a unique observation ID for submissions accepted from the MPC. The processing system automatically extracts these observation IDs from acknowledgment emails and uses them to perform routine queries through the MPC's WAMO service\footnote{\url{https://www.minorplanetcenter.net/wamo/}}. This process retrieves the latest ``observation status'' and matches the unique observation ID from the MPC with the unique \textit{diaSourceId} of a detection in our database. In addition, an orbit database is established to systematically record and manage the orbital information of the SSOs. This database is regularly updated using the \textit{MPCORB.DAT} file, which is refreshed daily on the MPC website. The orbit database ensures that we have the most current and accurate orbital parameters for known SSOs. In addition, an identification database is created to address cases in which some asteroids have multiple or different designations.

\section{Results} \label{sec:results}

From March 1, 2024 to May 31, 2024, the SSO processing system was applied in real time to the observational data collected during the WFST pilot survey. In this section, we present the application of the processing system, its performance metrics, and statistical results.

\subsection{Processing system applications and performances}

Following the completion of data processing by the WFST processing system each morning, the alert products generated from the difference images collected throughout the night are automatically transferred to our processing system. As described in Section \ref{sec:SSO Processing System}, the processing system begins with preprocessing steps, including \textit{avro} unpacking and formatting, known asteroids matching using \texttt{astcheck}, and false detection removal using machine learning.

\begin{figure}
    \centering
    \includegraphics[width=0.95\linewidth]{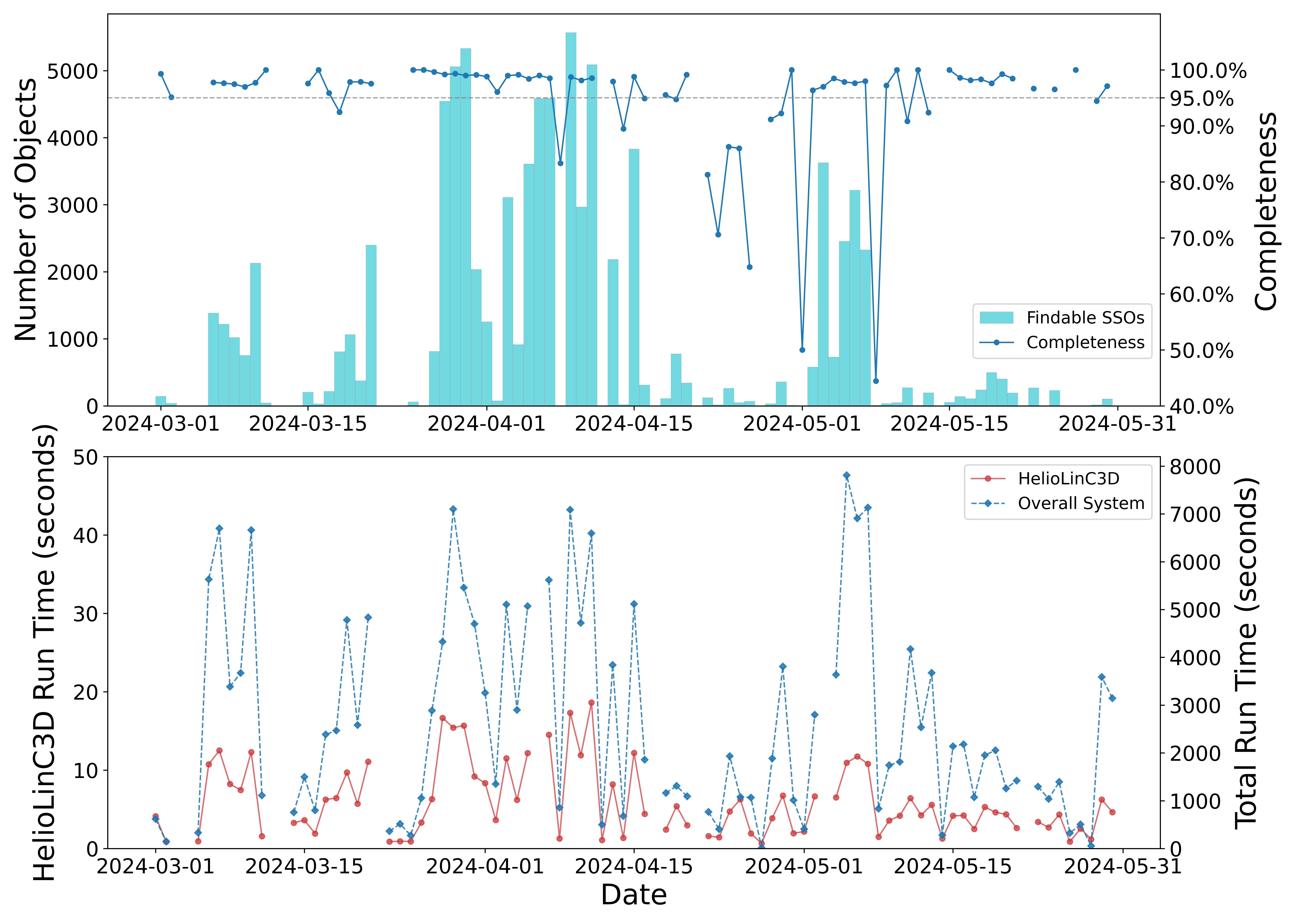}
    \caption{Completeness and run times of the processing system in single-night processing. Top: the number of findable SSOs and completeness of the processing system over dates. The gray line is 95.0\% completeness. Bottom: run time of the \texttt{HelioLinC3D}'s \textit{make\_tracklets} (red dot) and the entire processing system over dates (blue diamond).}
    \label{fig:one_night_pf}
\end{figure}

\begin{figure}
    \centering
    \includegraphics[width=0.95\linewidth]{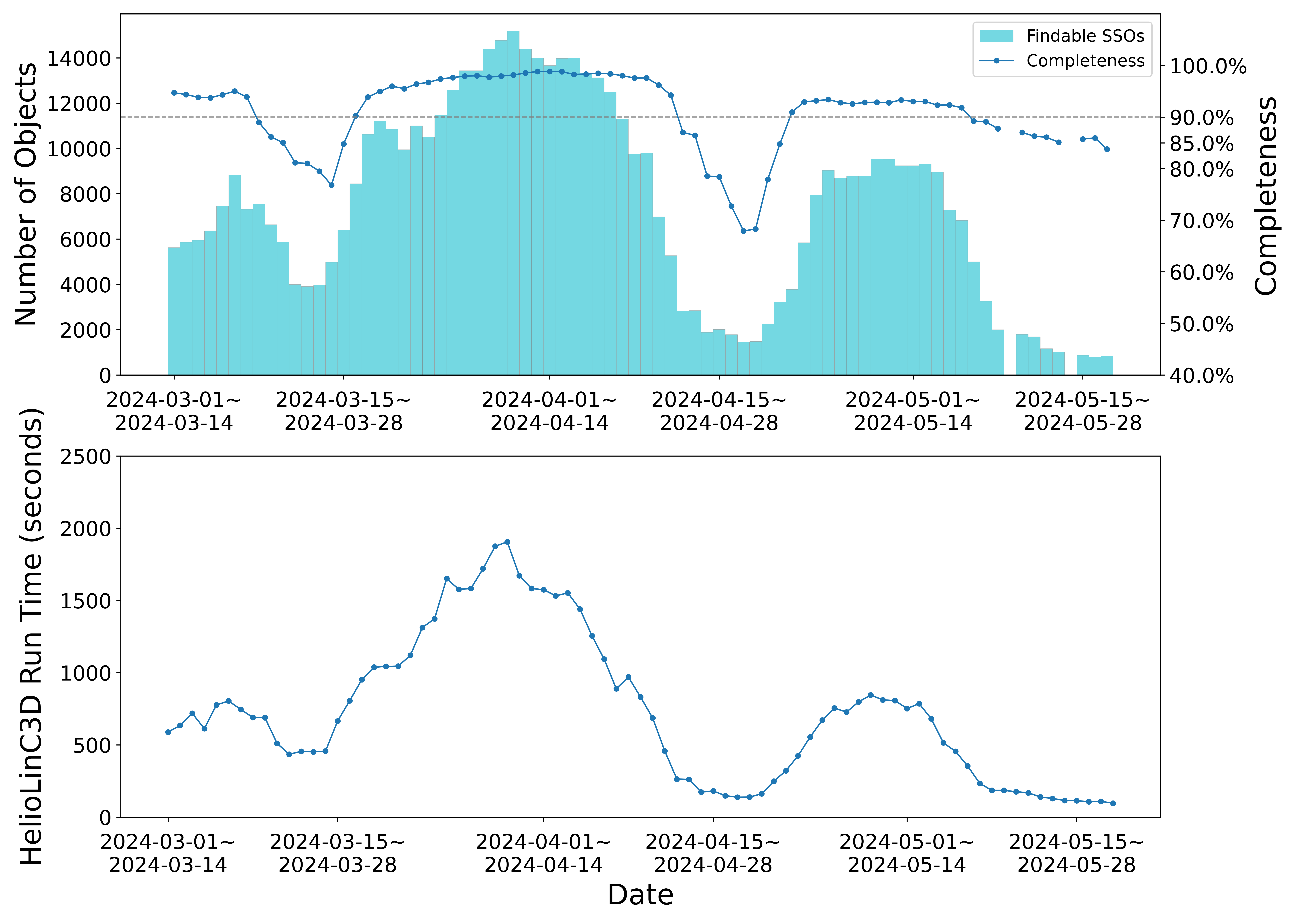}
    \caption{Completeness and run times of the processing system in multi-night processing. Top: the number of findable SSOs and the completeness of the processing system over dates. The gray line is 90.0\% completeness. Bottom: run times of the \texttt{HelioLinC3D} over dates.}
    \label{fig:multi_night_pf}
\end{figure}

To search SSO candidates, the processing system performs a single-night and a multi-night processing, with the results analyzed using \texttt{difi}. In single-night processing, most dates achieve a completeness rate exceeding 95\%. An object is classified as findable if it has three or more detections. The run time of the processing system for all nights is less than 8000 seconds, and each execution of \texttt{HelioLinC3D} reliably completes in less than 20 seconds, as illustrated in Figure \ref{fig:one_night_pf}. In multi-night processing, an object is considered to be findable if it has at least two detections on each of no fewer than three separate nights. The total processing time for each night in multi-night processing is less than 2000 seconds (see Figure \ref{fig:multi_night_pf}). Since the input data for multi-night processing have already been preprocessed during single-night processing, the majority of the time is spent on executing \texttt{HelioLinC3D}, while a lesser fraction is devoted to data reading and writing operations. Most dates achieve a completeness rate that exceeds 90\%.

Subsequently, the processing results, including both known SSOs and SSO candidates, undergo inspections. Verification of known SSOs is a rapid process that typically takes only a few seconds. This efficiency comes from the straightforward method of verifying these objects using their magnitudes. For SSO candidates, the daily count can reach hundreds. Despite this volume, most candidates are confirmed as genuine, and the evaluation of each candidate's \textit{gif} animation generally requires only a few seconds.

In practical applications, although the false positive rate is maintained within an acceptable range during the false detection removal phase, the application of this model to real-world data still yields a considerable number of false positives. This phenomenon is ascribed to the exclusion of numerous potential conditions from the training samples. To mitigate this issue, an enhancement of the dataset used for machine learning is planned in future iterations.

\subsection{Alert data and preprocessed results} \label{sec:Alert Data }

During the period, the SSO processing system processes a total of 60,266,765 alert entries. After removing false detections, the number of preprocessed alert entries decreased to 1,993,506, representing approximately 3.31\% of the initial dataset. The \texttt{astcheck} matching process successfully identified a total of 888,286 alerts corresponding to 38,879 SSOs as confirmed SSOs.

\subsection{Known asteroids} \label{sec:Known asteroids}

\begin{figure}
    \centering
    \includegraphics[width=1\linewidth]{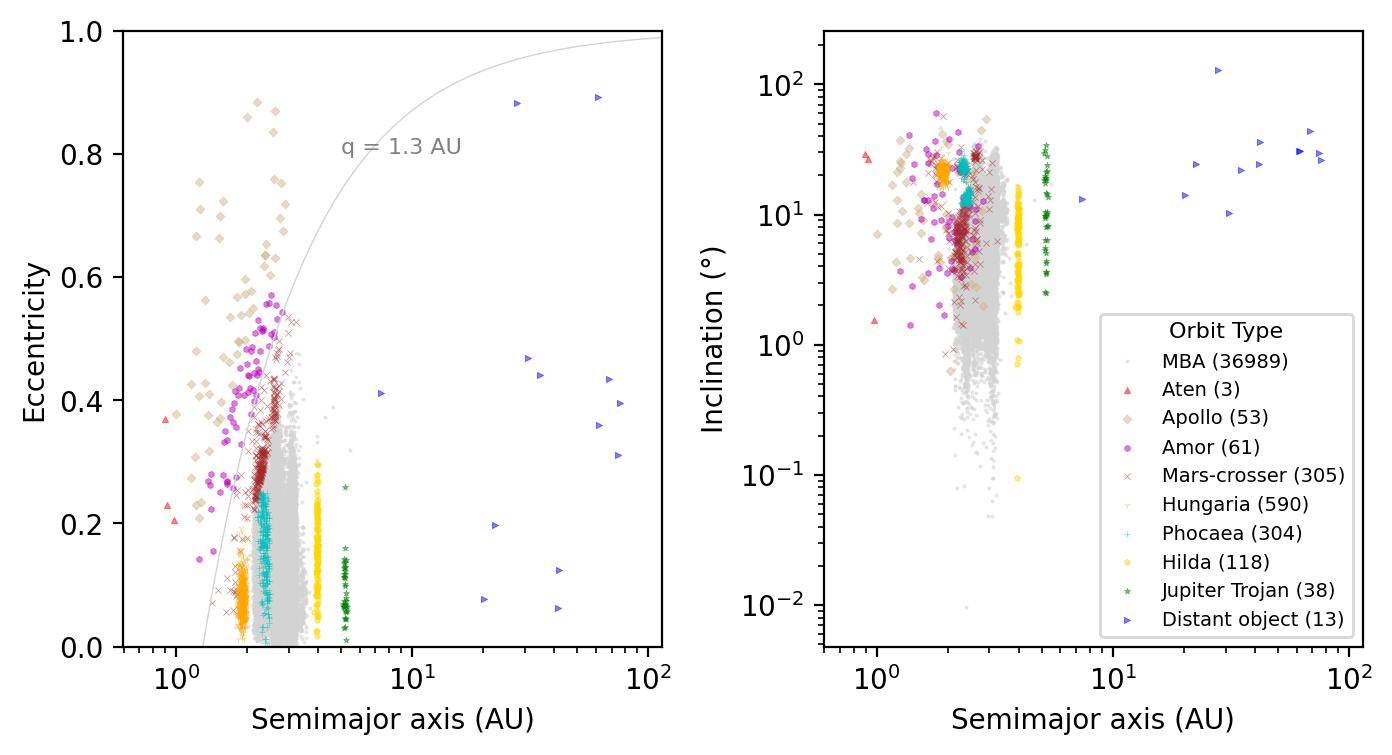}
    \caption{Orbital elements distribution of various types of 38,474 known asteroids observed by WFST, marked with different symbols. Left: the distribution of eccentricity versus semimajor axis, with a gray line indicating the orbital elements where q (perihelion distance) equals 1.3 AU. Right: the distribution of inclination versus semimajor axis.}
    \label{fig:known-asteroid}
\end{figure}

The wide FOV provided by the WFST, coupled with its robust survey capabilities, enables us to perform frequent repeated observations of known asteroids in a short time frame. During the period from March 1, 2024 to May 31, 2024, there were 658,489 observations of 38,520 known asteroids verified by the MPC in our submissions. Of the 38,520 identified known asteroids, the orbital distributions of 38,474 asteroids with various orbit types that are described in the MPC files are illustrated in Figure \ref{fig:known-asteroid}. The results of known SSOs recognized by the MPC after processing our submissions consist of: (1) known SSOs from step \ref{subsubsec:Automatic inspections for known SSOs} that we submitted directly and (2) a subset of SSO candidates from step \ref{subsubsec:Visual inspections for SSO candidates} that were initially not matched within our pipeline but were subsequently identified as known asteroids by the MPC.

\subsection{New asteroids} \label{sec:New asteroids}

\begin{figure}
    \centering
    \includegraphics[width=1\linewidth]{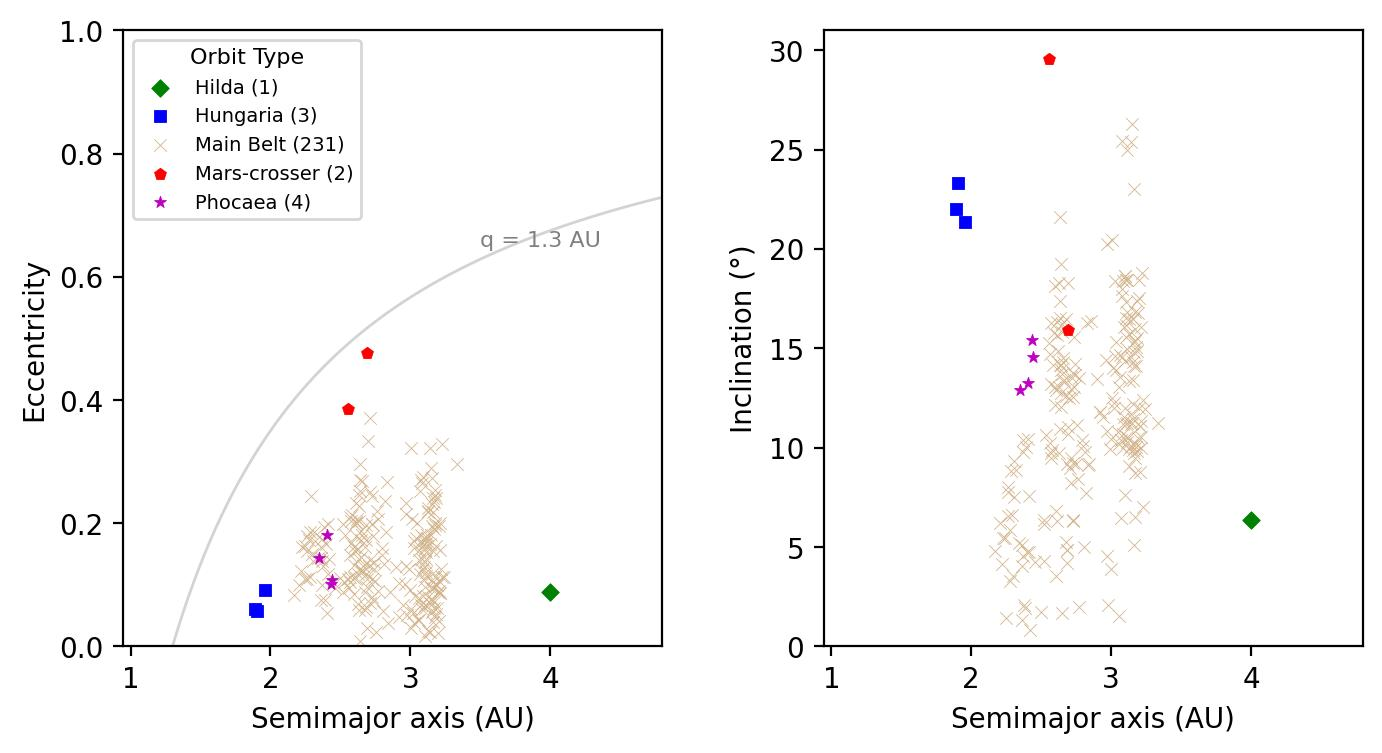}
    \caption{Orbital elements distribution of various types of SSO candidates initial reported by WFST, marked with different symbols. Left: the distribution of eccentricity versus semimajor axis, with a gray line indicating the orbital elements where q (perihelion distance) equals 1.3 AU. Right: the distribution of inclination versus semimajor axis.}
    \label{fig:orbits-wfst-asterisk-ustc}
\end{figure}

In the course of the WFST pilot survey, we submitted a cumulative total of 81 reports concerning SSO candidates to the MPC, covering 5,875 linkages and 24,036 observations. Upon verification by the MPC, 241 new asteroids received provisional designations and were initially reported by the WFST(see Figure \ref{fig:orbits-wfst-asterisk-ustc}). Furthermore, 2,125 observations were recorded in the ITF and are pending further observational data for linkage.

On the one hand, for single-night processing, there are 35 reports based solely on data from single-night observations, comprising 3,341 linkages with 11,456 observations. From these submissions, 24 new asteroids were confirmed by the MPC using only single-night observation data. On the other hand, the remaining 48 reports utilized data from multi-night observations within a sliding 14-day window, accounting for 2,534 linkages and 12,580 observations. In particular, 27\% of all newly identified asteroids confirmed by the MPC were discovered using a search method that comprised two observations per night over three different nights within the 14-day period (referred to as the $2 + 2 + 2$ mode). Furthermore, we find that multi-night linkages are more readily verified by the MPC compared to single-night tracklets, which require follow-up observations for confirmation. In this case, HelioLinC3D has been proven to be effective and efficient when applied to our wide field survey data.

\begin{deluxetable*}{ccccc}
\tablenum{2}
\tablecaption{WFST results on searching unknown asteroids in the pilot survey \label{tab:wfst_unknown_resluts}}
\tablewidth{0pt}
\tablehead{
\colhead{Observation Type} & \colhead{Number of Reports} & \colhead{Number of Observations} & \colhead{Number of Linkages} & \colhead{Number of New Asteroids }
}
\startdata
Single-night data & 35 & 11456 & 3341 & 24 \\
Multi-night data & 46 & 12580 & 2534 &  217\\
Total & 81 & 24036 & 5875 & 241 \\
\enddata
\end{deluxetable*}

\section{Summary}\label{sec:sum}

This work details the architecture, design, and processing flow of our SSO processing system applied to WFST, utilizing the innovative HelioLinC3D algorithm. The processing system is capable of identifying known asteroids and searching SSO candidates through both single-night and multi-night processing. Using data from a three-month pilot survey, we reported 658,489 observations of 38,520 objects of known asteroids and 24,036 observations of 5,875 linkages of SSO candidates. A total of 241 newly discovered asteroids received provisional designations from the MPC, while 2,404 observations were placed into the MPC's ITF for further confirmation. Notably, 27\% of the newly identified asteroids were identified exclusively in the multi-night processing that entails two observations each night for three nights within a 14-day period, referred to as the $2+2+2$ mode. Our results demonstrate that the processing system achieves significant performance in terms of time efficiency and completeness.

The results highlight the capabilities of the SSO processing system, which effectively handles both single-night and multi-night processing. The successful application of HelioLinC3D to WFST, particularly in multi-night processing, underscores its efficiency and productivity in the context of wide-field surveys. This is supported by extensive real-world observational data. Furthermore, WFST is emerging as a significant contributor to asteroid detection, showing considerable promise in advancing the study of the solar system. Its capabilities are expected to enhance our understanding of small solar system bodies and contribute valuable data for future research.

The processing system continuously processes results and undergoes ongoing adjustments to optimize performance. Future improvements will focus on addressing challenges such as poor photometry, improving data sets and machine learning models, resolving overflow errors, and other operational issues. 

Looking ahead, our goal is to use the accumulated data for deeper investigations into areas such as NEOs, light curves, and activity monitoring. Furthermore, we plan to integrate advanced tools such as Kernel-Based Moving Object Detection \citetext{KBMOD; \citealp{Whidden_2019, Smotherman_2021}}and adopt innovative methods to detect NEOs \citep{Ye_2019}. These enhancements are expected to significantly broaden the capabilities of WFST to facilitate discoveries across a wider spectrum of solar system studies.

\begin{acknowledgments}
The Wide Field Survey Telescope (WFST) is a joint facility of the University of Science and Technology of China, Purple Mountain Observatory. This work is supported by National Key Research and Development Program of China (2023YFA1608100). The authors gratefully acknowledge the support of the National Natural Science Foundation of China (NSFC, Grant Nos. 12173037, 12233008), the CAS Project for Young Scientists in Basic Research (No. YSBR-092), the Fundamental Research Funds for the Central Universities (WK3440000006) and Cyrus Chun Ying Tang Foundations.
\end{acknowledgments}

\vspace{5mm}
\facilities{WFST:2.5m}

\software{\texttt{astropy} \citep{astropy:2013, astropy:2018, astropy:2022}, \texttt{Jupyter} \citep{2007CSE.....9c..21P, kluyver2016jupyter}, \texttt{matplotlib} \citep{Hunter:2007}, \texttt{numpy} \citep{numpy}, \texttt{pandas} \citep{mckinney-proc-scipy-2010, pandas_10957263}, \texttt{python} \citep{python}, \texttt{scipy} \citep{2020SciPy-NMeth, scipy_11255513}, \texttt{HelioLinC3D} \citep{holman2018}, \texttt{difi} \citep{difi}, \texttt{astcheck} \citep{astcheck}, \texttt{braai} \citep{Duev_2019} and \texttt{mongodb} \citep{mongodb}}

\bibliography{ms}{}
\bibliographystyle{aasjournal}

\end{CJK*}
\end{document}